\documentclass{aa}  
\usepackage{natbib}
\bibpunct{(}{)}{;}{a}{}{,}
\usepackage{graphicx}
\usepackage{txfonts}
\begin{document}
 \title{
Submillimeter Line Emission from LMC N159W: a Dense, Clumpy PDR in a Low
Metallicity Environment
}
\titlerunning{Submillimeter Line Emission from LMC N159W}
\authorrunning{Pineda, Mizuno, Stutzki, Cubick et al.}

   \author{   
     J.L.\,Pineda\inst{1} \and
     N.\,Mizuno\inst{2} \and
 J.\,Stutzki\inst{3} \and
 M.\,Cubick\inst{3} \and
     M.\,Aravena\inst{1} \and
     F.\,Bensch\inst{1} \and
     F.\,Bertoldi\inst{1} \and
     L.\,Bronfman\inst{4} \and
     K.\,Fujishita\inst{2} \and
     U.U.\,Graf\inst{3} \and
     M.\,Hitschfeld\inst{3} \and
     N.\,Honingh\inst{3} \and
     H.\,Jakob\inst{3} \and
     K.\,Jacobs\inst{3} \and
     A.\,Kawamura\inst{2} \and
     U.\,Klein\inst{1} \and 
     C.\,Kramer\inst{3} \and
    J.\,May\inst{4} \and
     M.\,Miller\inst{3} \and
    Y.\,Mizuno\inst{2} \and
     P.\,M\"uller\inst{1} \and
     T.\,Onishi\inst{2} \and
     V.\,Ossenkopf\inst{3} \and
     D.\,Rabanus\inst{3} \and
     M.\,R\"ollig\inst{1} \and
     M.\,Rubio\inst{4} \and
     H.\,Sasago\inst{2} \and
     R.\,Schieder\inst{3} \and
     R.\,Simon\inst{3} \and
     K.\,Sun\inst{3} \and
     N.\,Volgenau\inst{3} \and 
     H.\,Yamamoto\inst{2} \and 
   Y.\,Fukui\inst{2} 
          }   

 \offprints{J.\,L.\,Pineda \email{jopineda@astro.uni-bonn.de}}

   \institute{   
  Argelander-Institut f\"ur Astronomie,  Auf dem H\"ugel 71,
     D-53121 Bonn, Germany
     \and
     Department of Astrophysics, Nagoya University, Chikusa-ku, Nagoya 464-8602, Japan
     \and
     KOSMA, I. Physikalisches Institut, Universit\"at zu K\"oln,   
     Z\"ulpicher Stra\ss{}e 77, D-50937 K\"oln, Germany    
     \and
     Departamento de Astronom\'{i}a, Universidad de Chile, Casilla 36-D, Santiago, Chile
   }

\date{Received / Accepted } 
   \abstract 
     { 
 Star formation at earlier cosmological times takes place in an
     interstellar medium with low metallicity. The Large Magellanic Cloud 
     (LMC) is ideally suited to study star formation in such an environment. }
     { 
     The physical and chemical state of the ISM in a star forming environment
     can be constrained by observations of submm and FIR spectral lines of the 
     main
     carbon carrying species, CO, C\,{\sc i} and C\,{\sc ii}, which originate
     in the surface layers of molecular clouds illuminated by the UV radiation
     of the newly formed, young stars. }
     { 
       We present high-angular resolution sub-millimeter observations in the
       N159W region in the LMC obtained with the NANTEN2 telescope of the
       $^{12}$CO $J = 4 \to3$, $J = 7\to6$, and $^{13}$CO $ J = 4 \to 3$
       rotational and [C\,{\sc i}] $^3$P$_1-^3$P$_0$ and $^3$P$_2-^3$P$_1$
       fine-structure transitions.  The $^{13}$CO $J =4 \to 3$ and [C\,{\sc
         i}] $^3$P$_2-^3$P$_1$ transitions are detected for the first time in
       the LMC. We derive the physical and chemical properties of the
       low-metallicity molecular gas using an escape probability code and a
       self-consistent solution of the chemistry and thermal balance of the
       gas in the framework of a clumpy cloud PDR model.}
     { 
       The separate excitation analysis of the submm CO lines and the carbon
       fine structure lines shows that the emitting gas in the N159W region
       has temperatures of about 80 K and densities of about
       10$^{4}$cm$^{-3}$.  The estimated C to CO abundance ratio close to
       unity is substantially higher than  in dense massive star-forming
         regions  in the Milky Way.  The analysis of all observed lines
       together, including the [C\,{\sc ii}] line intensity reported in the
       literature, in the context of a clumpy cloud PDR model constrains the
       UV intensity to about $\chi \approx 220$ and an average density of
         the clump ensemble of about 10$^5$ cm$^{-3}$,  thus confirming
         the presence of high density material in the LMC N159W region.  }{}

 \maketitle
%

\section{Introduction}

Since the formation of the first generation of stars in our Universe, the
cyclic process of formation and destruction of stars has progressively
enriched the interstellar medium (ISM) with heavy elements and dust. This
implies that, at earlier cosmological times, stars formed in lower
metallicity, lower dust-to-gas ratio molecular gas. Newly formed massive stars
illuminate their progenitor molecular clouds with far-ultraviolet photons
(FUV; 6\,eV $<$ h$\nu< 13.6$\,eV) producing photon-dominated regions (PDRs;
\citealt[][and references therein]{HollenbachTielens99}). The lower
dust-to-gas ratio reduces the ability of the low-metallicity molecular cloud
to attenuate FUV photons, making low-metallicity PDRs more extended than their
Solar metallicity counterparts.  This has an impact on the C$^{+}$/C/CO
transition layer in PDRs, as CO is more efficiently photo-dissociated
\citep{vanDishBlack88}, enhancing the C$^{+}$ and C abundance relative to CO.
Therefore, star formation in low metallicity gas differs from that in solar
metallicity gas, as the structure and thermal balance of the progenitor
molecular gas is different.  The extent to which the properties of the
molecular gas are modified have been theoretically studied
\citep[e.g.][]{MaloneyBlack88,Lequeux94,Bolatto99,Kaufman99,Roellig06} and to
some extent observationally constrained \citep[e.g.][]{Israel97,Madden06}.
The study of the properties of the low metallicity molecular gas is of great
importance in order to understand the evolution of youngest galaxies in our
Universe. 

Nearby dwarf and irregular galaxies are examples where we can observe massive
star formation taking place in low-metallicity molecular gas.  One of the best
studied examples is the Large Magellanic Cloud (LMC).  Due to its proximity
\citep[50\,kpc;][]{Feast99} and nearly face-on orientation \citep[$i =
35\degr$;][]{vanderMarel01}, the LMC is ideal to study the impact of massive
star formation in its low-metallicity (Z $\sim$ 0.3$-$0.5 Z$_\odot$;
\citealt{Westerlund97}) molecular gas.  The LMC has been entirely mapped in
[C\,{\sc ii}] 158\,$\mu$m by \citet{Mochizuki94} and in $^{12}$CO $J = 1 \to
0$ by \citet{Cohen88} at low resolution of 8.8 arcmin.  \citet{Fukui99}
revealed the first spatially resolved image of the giant molecular clouds over
the LMC in the $^{12}$CO $J =1 \to 0$ emission at 2.6 arcmin resolution
($\sim$ 40\,pc) by using the NANTEN 4m telescope.  The subsequent CO studies
with NANTEN offer the details of the molecular and star-formation properties
of these GMCs in the LMC
\citep{Mizuno01,Yamaguchi2001,Blitz2007,Fukui2002,Fukui2007}. The CO
observations of the LMC show that the molecular gas towards the molecular
cloud complex associated with the N159 H\,{\sc ii} region \citep{Henize56} is
an outstanding star forming region which is part of the most significant
molecular concentration in the galaxy.  An extensive study of the CO
  emission with improved angular resolution has been carried out within the
  ESO-SEST Key Programme on CO in the Magellanic Clouds (\citealt{Israel93}
  and subsequent publications).

Observations of low$-J$ $^{12}$CO and $^{13}$CO transitions with improved
angular resolution ($\sim$10\,pc) have unveiled three giant molecular clouds
(GMCs); N159E, N159S, and N159W \citep{Johansson98,Bolatto00}. At similar
scales, the [C\,{\sc ii}] distribution in the N159 complex has been mapped by
\citet{Boreiko91} and \citet{Israel96}.  At lower angular resolution of
110\arcsec, and hence averaging over larger spatial scales of about 50 pc, the
[C\,{\sc i}] $^3$P$_1-^3$P$_0$ and $^{12}$CO $J = 4 \to 3$ transitions have
been detected in this region with the AST/RO telescope
\citep{Stark97,Bolatto00}.

The brightest CO peak in the entire LMC is N159W.  In this region, an
excitation analysis made by \citet{Johansson98}, based in low$-J$ $^{12}$CO
and $^{13}$CO transitions, suggested gas densities of $n = 10^4$\,cm$^{-3}$
and $T_\mathrm{kin} = 10 - 20$\,K.  A similar analysis by
\citet{Minamidani2007}, including the $^{12}$CO $J = 3 \to 2$ transition,
resulted in similar gas densities but somewhat higher temperatures
($T_\mathrm{kin}>30$\,K).  Including $^{12}$CO $J = 4 \to 3$ observations,
\citet{Bolatto05} argue that the CO intensities are best explained by a warm
($T_\mathrm{kin}=100$\,K) and low-density ($n=$10$^2$\,cm$^{-3}$) and a cold
($T_\mathrm{kin}=20$\,K) and dense ($n=$10$^5$\,cm$^{-3}$) components. Note
that the work of \citet{Johansson98} and \citet{Bolatto05} apply to different
spatial scales (10\,pc and 26\,pc, respectively).

The PDRs in N159W have been studied by \citet{Israel96}, \citet{Bolatto00},
and \citet{Pak1998}.  Based on different tracers of the strength of the
incident FUV radiation field, \citet{Israel96} derived a value between $\chi =
120-350$\footnote{\citet{Israel96} originally derived this value in units of
  the \citet{Habing68} field. The \citet{Draine78} field is related to the
  \citet{Habing68} field by a factor of 1.71 when averaged over the full FUV
  range.} (in units of the intensity of the mean interstellar radiation field
derived by \citealt{Draine78}).  \citet{Bolatto00} found that the
$I_\mathrm{[C\,{\sc II}]}/I_\mathrm{CO}$ ratio is enhanced compared to that in
similar Galactic star-forming regions while the $I_\mathrm{[C\,{\sc
    I}]}/I_\mathrm{CO}$ remains similar.  This is in agreement with a simple
analytical model by \citet{Bolatto99} that predicts growth of the [C\,{\sc
  ii}] and [C\,{\sc i}] emitting layers with decreasing metallicity, resulting
from a shrinking of the CO molecular core.  The N159W region was included in
PDR model calculations by \citet{Pak1998}, which constrained the physical
conditions of the line-emitting gas to 10$^{3.7}$ cm$^{-3} < n < 10^{4.7}$
cm$^{-3}$ and 60 $< \chi <$ 600.

In this paper, we present high-resolution (10\,pc) observations of the N159W
region obtained with the new NANTEN2 telescope. We present detections of the
$^{12}$CO $J = 4 \to 3$, $J = 7 \to 6$, and $^{13}$CO $J = 4 \to 3$ rotational
and [C\,{\sc i}] $^3$P$_1-^3$P$_0$ and $^3$P$_2-^3$P$_1$ fine-structure
transitions.  The $^{13}$CO $J = 4 \to 3$ and [C\,{\sc i}] $^3$P$_2-^3$P$_1$
transitions are detected for the first time in the LMC.  The $^{12}$CO $J = 7
\to 6$ line has been detected at the low angular resolution of the AST/RO
telescope towards the 30 Dor region \citep{Kim2006}.  The NANTEN2 observations
are described in Section 2.  We derive physical properties of the
low-metallicity molecular gas using an escape probability radiative transfer
code independently for the CO and [C\,{\sc i}] lines (Section 3), and a
self-consistent solution of the chemistry and thermal balance of the gas is
given using a photon-dominated region (PDR) model for a clumpy interstellar
medium (Section 4), explaining the observed emission of all observed lines,
including low$-J$ CO lines and [C\,{\sc ii}] 158 $\mu$m from the literature.
We discuss our results in Section 5, and summarize in Section 6.

\section {Observations}

We use the new NANTEN2 4m telescope situated at 4865\,m altitude at Pampa la
Bola in northern Chile to observe the $^{12}$CO $J = 4 \to 3$ (461.0408\,GHz)
, $J = 7 \to 6$ (806.6517\,GHz), and $^{13}$CO $J = 4 \to 3$ (440.7654\,GHz)
rotational and [C\,{\sc i}] $^3$P$_1-^3$P$_0$ (492.1607\,GHz) and
$^3$P$_2-^3$P$_1$ (809.3446\,GHz) fine-structure transitions toward N159W. The
line parameters derived from the NANTEN2 observations of LMC N159W are listed
in  Table~\ref{tab:gauss}.

The observations were made toward the peak $^{12}$CO $J = 1 \to 0$
intensity position in N159W, located at $\alpha$ = 5$^{\tiny
\textrm{h}}$39$^{\tiny \textrm{m}}$36\fs8 and $\delta$ =
$-$69\degr45\arcmin31\farcs9 (J2000), using the total-power observing
mode.  We use a reference position at $\alpha$ = 5$^{\tiny
\textrm{h}}$38$^{\tiny \textrm{m}}$53\fs2 and $\delta$ =
$-$69\degr46\arcmin28\farcs8 (J2000), which is free of $^{12}$CO $J =
1 \to 0$ emission (3 $\sigma$ upper limit of 6.2 K km\,s$^{-1}$).  The
duration of each beam switch cycle was varied between 20 and 30
seconds, depending on the atmospheric stability during the
observations. The final spectra result from total integration times of
20 mins to about 1 hour.  The pointing was checked regularly on
Jupiter, IRC+10216, and IRc2 in OrionA.  The applied corrections were
always smaller than 20\arcsec, and usually less than 10\arcsec.

The observations were conducted with a dual-channel 460/810\,GHz receiver
installed for verifying the submillimiter performance of the telescope.
Double-sideband (DSB) receiver temperatures were $\sim$ 250\,K in the lower
channel and $\sim$ 750\,K in the upper one. The intermediate frequencies (IF)
are 4\,GHz and 1.5\,GHz, respectively. The latter IF allows simultaneous
observations of the $^{12}$CO $J = 7 \to 6$ line in the lower and of the
[C\,{\sc i}] $^3$P$_2-^3$P$_1$ line in the upper sideband. These two lines are
observed simultaneously with one of the lines in the 460\,GHz channel. As
backends we used two acousto-optical spectrometers (AOS) with a bandwidth of
1\,GHz and a channel resolution of 0.37\,km s$^{-1}$ at 460\,GHz and 0.21\,km
s$^{-1}$ at 806\,GHz.

The LMC observations with NANTEN2 were performed in several sessions during
the September to November 2006 period. In all sessions both channels of the
dual frequency test-receiver were active, so that the determination of the
atmospheric transmission for both channels can be done in a single step with a
common atmospheric model. This new atmospheric calibration scheme, implemented
as part of the standard observing software {\it kosma\_control} developed at
Universit\"at zu K\"oln and installed at NANTEN2, uses a parametrization of
the \cite{Pardo2004} atmospheric model ATM.  The single, free parameter,
fitted to the observed atmospheric emission spectrum derived from
hot/cold/sky-measurements in the standard calibration cycle, is the
precipitable water vapor $pwv$. The parameterized model provides frequency
lookup tables of the dry ($a_\nu$) and wet ($b_\nu$) contribution to the
opacity of the atmosphere above the observing site, $\tau_\nu = a_\nu + b_\nu
~ pwv $.  The $a_\nu$ and $b_\nu$ -coefficients are derived from the ATM model
atmosphere calculated for the 4865~m altitude of the NANTEN2 observatory.  The
cold-load is a liquid nitrogen load, the hot-load is at the cabin temperature,
which is monitored electronically. The sky temperature is taken as the ambient
outside temperature monitored by the weather station of the observatory. This
assumption is occasionally verified through sky-dip measurements, which are
also used to determine the forward/spillover efficiency.  Using this
calibration, the relative intensities of lines observed in different receiver
bands simultaneously is, beyond the signal/noise of the data, limited only by
the precision and validity of the atmospheric model. The main advantage of
this new scheme is that it gives a much higher precision for the determination
of the atmospheric line-of-sight transmission in receiver bands with low
transmission and poorer sensitivity, because it includes the stronger
(hot-sky) difference signal of the higher transmission bands and the higher
signal-to-noise data from the more sensitive bands in determining the
transmission at each frequency point.  It also fully takes into account
frequency dependent variations of the atmospheric opacity across the receiver
bands, such as resulting from narrow atmospheric lines and tails of wide ones.
Comparison with the traditional calibration approach which determines the
atmospheric transmission as an average over a representative section of each
receiver band individually from hot/cold/sky-measurements, shows a much higher
robustness against errors in the values for the ambient and effective
atmospheric temperatures used in the calibration.

The other factors that enter in the relative calibration of the intensity of
lines from different receiver bands are the ratio of their main beam
efficiency and the pickup from the error beam pattern outside the main beam.
Main beam efficiencies were determined from Jupiter radio continuum scans
\citep{Simon2007}, assuming a brightness temperature of Jupiter of 154 K and
126 K respectively at 460-490 and 810 GHz \citep{Griffin1986}.  Comparison of
the fitted angular width of the Jupiter scan with the disk diameter of Jupiter
allows to derive the FWHM of the main beam, and from its brightness, the main
beam efficiency.  The thus derived observational parameters are listed in
Table~\ref{tab:obsparms}.  A full description of this new calibration scheme
for ground based submm observations is in preparation for a separate,
technical paper.

A coarse estimate of the error beam of the NANTEN2 telescope was obtained by
analyzing scans across the edge of the Moon and the Sun \citep{Simon2007}.
These show a signal consistent with an error beam size of some 400\arcsec\,at
460 GHz and 240\arcsec\,at 810 GHz, consistent with the diffraction pattern of
individual panels, which have a size of 50 by 70 cm, and hence due to panel
mis-alignment.  We estimate the integrated power (relative to the main beam)
of these error beams to be 10\% and 15\% at 460-490 GHz, respectively 810 GHz.
The error beam intensities are consistent with a surface rms due to panel
misalignment of about 25 $\mu$m, as independently confirmed by photogrammetric
measurements.  From the analysis of these data, we estimate the ratio of main
beam efficiencies to be determined to a precision of about 10\%, resulting in
a similar systematic error in the observed intensity ratios of lines in the
460 and 810 GHz band.

The simultaneous coverage of both the $^{12}$CO $J = 7 \to 6$ and [C\,{\sc i}]
$^{3}$P$_{2} \to ^{3}$P$_{1}$ transitions in the upper and lower sideband
implies that their relative calibration, including the atmospheric imbalance
between the two sidebands, is very precise, limited only by the radiometric
S/N in this receiver channel.  The lower frequency channel of the dual channel
receiver was alternatingly tuned to the $^{12}$CO $ J = 4 \to 3$, the [C\,{\sc
  i}] $^{3}$P$_{1} - ^{3}$P$_{0}$ transition, and the $^{13}$CO $J = 4 \to 3$
line. The atmospheric transmission was particularly good for an initial set of
observations covering the $^{12}$CO $J = 4 \to 3$ line in the lower channel.
At this time the zenith opacity was about 0.40 at the $^{12}$CO $J = 4 \to 3$
line frequency and about 0.92 for the $^{12}$CO $J = 7 \to 6$ line, resulting
in very high S/N spectra in particular for the $^{12}$CO $J = 4 \to 3$ line.
Using the $a_\nu$ and $b_\nu$ coefficients from the \citet{Pardo2004} model
these zenith optical depths correspond to a precipitable water vapor (pwv) of
about 0.51 mm.  A subsequent observing session covering the [C\,{\sc i}]
$^{3}$P$_{1} - ^{3}$P$_{0}$ transition had again a very good zenith opacity of
about 0.5 to 0.6 at 492 GHz, and about 0.7 to 0.8 in the upper frequency
channel for the CO $7\to 6$ line (pwv $\simeq$ 0.38 to 0.44 mm).  During the
$^{13}$CO $J = 4 \to 3$ line observations the weather was very good for a
short session, with an opacity of about 0.5 to 0.6 at the line frequency of
440 GHz, at the edge of the 600 $\mu$m atmospheric window, and correspondingly
low zenith opacity of about 0.6 at the $^{12}$CO $J = 7 \to 6$ frequency (pwv
$\simeq$ 0.31 mm). Most of the $^{13}$CO $J = 4 \to 3$ data were taken in a
long integration at substantially higher zenith opacity, varying around
1.2-1.5 (corresponding to about 1.4 to 1.5,  pwv $\simeq$ 0.70 to 0.89
  mm, at 806 GHz).  While this implies that the absolute calibration at these
high opacities may easily be influenced by systematic errors (such as
non-perfect temperature estimates for the effective radiative temperature of
the atmosphere), the relative calibration between the low- and high-frequency
line is largely unaffected and thus rather accurate.

In the following, we thus used estimates of the systematic errors for the line
intensity ratios introduced through the calibration uncertainty (as explained,
these are mainly due to the precision with which the beam efficiency ratio is
determined), as quoted in Table~\ref{tab:gauss_corrected} and add these
quadratically with the radiometric noise, assumed to be given by the formal
errors of the Gaussian fits.

\section{Observational Results}

\subsection{NANTEN2 data}

\begin{table*} [t]
\caption{Gaussian-fit line parameters derived from the NANTEN2 observations of
  LMC N159W and from the literature}
\label{tab:gauss}
\centering
\begin{tabular}{c c c c c}
\hline\hline
Line & amplitude$^{(a)}$  & center$^{(b)}$ & FWHM$^{(b)}$ & integral \\
     &  [K]  &  [km\,s$^{-1}$] & [km\,s$^{-1}$]&[K\,km\,s$^{-1}$]  \\
\hline
\\[-0.2cm]
\multicolumn {5}{l}{NANTEN2 data}
\\
$^{12}$CO $J = 4 \to 3$ & 9.31 (0.05) & 238.3 (0.016) & 7.32 (0.04) & 72.6 (0.3) \\
$^{12}$CO $J = 7 \to 6$ & 3.33 (0.09) & \\
$^{13}$CO $J = 4 \to 3$ & 0.95 (0.10) \\

[C\,{\sc i}] $^3$P$_1\to ^3$P$_0$  & 1.51 (0.05) \\ 

[C\,{\sc i}] $^3$P$_2\to ^3$P$_1$  & 1.87 (0.08) \\
\hline
\multicolumn {5}{l}{literature data$^{(c)}$}\\
$^{12}$CO $J = 1 \to 0$ & 6.79 (0.15) & 238.0 (0.073) & 6.95 (0.17) & 50.0 (1.6) \\
$^{13}$CO $J = 1 \to 0$ & 0.78 (0.08) & 238.0 (0.35) & 6.81 (0.83) & 5.6 (0.9)\\

[C\,{\sc ii}] & 7.7 (0.9) & 232.8 (0.8) & 9.8 (0.9) & 75.5 (7.0)\\
\hline

\multicolumn {5}{l}{$^{(a)}$  amplitude derived from fitted Gaussian width and integral} \\

\multicolumn {5}{l}{$^{(b)}$ center and width fixed to $^{12}$CO $J = 4 \to 3$
  best fit values }\\
\multicolumn {5}{l} {$^{(c)}$ references for literature data see Section 3.2}
\end{tabular}
\end{table*}

\begin{figure}[t]
  \centering
  \includegraphics[width=0.45\textwidth,angle=0]{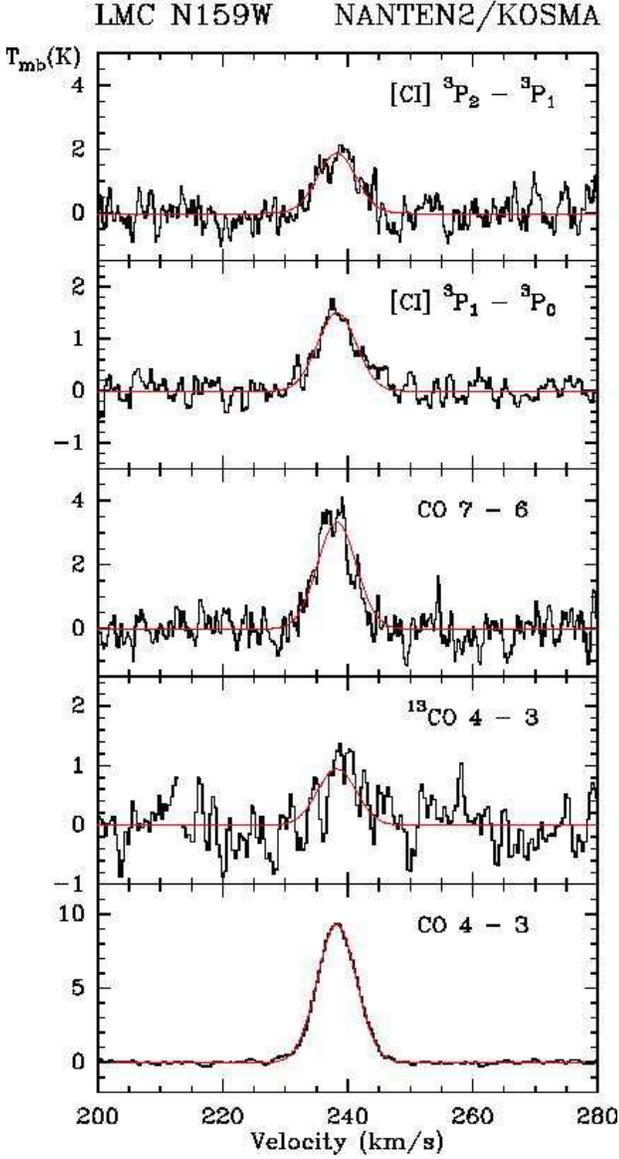}
     \caption{LMC N159W: observed lines and Gaussian fits 
(red line).
}
\label{fig:plot_all_spectra}
\end{figure}

\begin{table*} [t]
\caption{NANTEN2/dual-channel test receiver: Observational Parameters}
\label{tab:obsparms}
\centering
\begin{tabular}{c c c}
\hline\hline
receiver channel & 460-490 GHz & 810 GHz \\
\hline
\\[-0.2cm]
$\eta_{mb}$  &    0.5 & 0.4 \\
beam FWHM [\arcsec ] & 38 & 26 \\
$T_{rec}$ [K] &  250 & 750 \\
\hline\\
\end{tabular}
\end{table*}

The spectra observed towards LMC N159W are shown in
Fig.~\ref{fig:plot_all_spectra}, together with Gaussian fits of the center,
width and integrated line intensity, resp.\ amplitude. The results of the
Gaussian fits are listed in Table~\ref{tab:gauss}. The line center and width
determined from the high signal-to-noise $^{12}$CO $J = 4 \to 3$ spectrum were
taken and held fixed for the Gaussian fit of the line amplitude of the other
lines. This is adequate as their profiles are, within their limited
signal-to-noise, consistent with that of the $^{12}$CO $J = 4 \to 3$ line.

As explained above, due to the simultaneous observations and the joined
atmospheric transmission calibration for the lines in the upper and lower
frequency channel, the intensity ratio of the lines observed is rather
precisely determined and is mainly limited by the estimated systematic errors
due to the determination of the beam efficiencies as discussed above.  The
only further correction concerns the different beam size of the low and high
frequency channel of 38\arcsec, resp.\ 26\arcsec.  Unfortunately, we did not
have enough nights with superb weather to make small maps around the peak
positions at (0,0) in order to spatially smooth the high frequency
observations to the angular resolution of the low frequency beam.  For
correcting the beam size effects for the high frequency observations we thus
have to rely on an independent determination of the source extent. AST/RO has
repeatedly observed the $^{12}$CO $J = 4 \to 3$ line, once with a relatively
large beam FWHM of 204\arcsec\,\citep{Bolatto00}, resulting from improper
alignment of the AST/RO optical components, giving an integrated intensity of
8.4 K km\,s$^{-1}$, and with the diffraction beam size of the AST/RO telescope
at 460 GHz of 109\arcsec \citep{Bolatto05}, giving an integrated line
intensity of 20.8 K km\,s$^{-1}$. Assuming a Gaussian source distribution with
FWHM $\Theta_s$ and a source peak brightness temperature of $T_{s,peak}$, the
beam filling correction gives a main beam brightness temperature in a beam of
FWHM $\Theta_b$ of $ T_{mb}={\Theta_s^2}/{(\Theta_s^2+\Theta_b^2}) T_{s,peak}$
.  The ratio of the main beam brightness temperature in two different beams 1
and 2 is given by $R_{1,2} = {T_{mb,1}}/ {T_{mb,2}} = {(\Theta_{b,2}^2 +
  \Theta_s^2)} /{(\Theta_{b,1}^2+\Theta_s^2)} $.  With this, we can derive the
source size from the observed brightness ratio in two beam sizes:
\begin{displaymath}
\Theta_s^2 = \frac {R_{1,2} \Theta_{b,1}^2 - \Theta_{b,2}^2} {1-R_{1,2}},
\end{displaymath}
giving $\Theta_s = 52''$ from comparing the NANTEN2 $^{12}$CO $J = 4 \to 3$
data with the early, and $\Theta_s = 61''$ for comparison with the later
AST/RO data.

We thus adopt a source size of 60\arcsec\,and correct the observed intensity
ratios for the beam coupling of the different beam sizes at two observing
bands $x$ and $y$ to the source intrinsic intensity ratio $R_s$ via

\begin{displaymath}
R_s = \frac {T_{mb,x}}
{T_{mb,y}
} \, \frac {\Theta_s^2+\Theta_{b,x}^2}
{\Theta_s^2+\Theta_{b,y}^2} .
\end{displaymath}

This correction assumes that the source structure is the same at the spatial
scales observed with the 460 and 810 GHz channels.  The validity of this
assumption likely dominates the uncertainties in the line ratios.  However, as
we will see in Section~\ref{sec:escape-prob-radi} and
\ref{sec:pdr-model-analysis}, the good fit obtained for the physical
parameters supports this assumption.  Table~\ref{tab:gauss_corrected} shows
the line ratios thus corrected and used for the excitation analysis discussed
below.

\subsection{Additional Lines from the Literature }

Beyond the lines newly observed with NANTEN2 we have included in the analysis
described in the following also the $^{12}$CO and $^{13}$CO $J=1\to 0$ lines
(the $^{12}$CO data are presented in \citet{Pineda08b} and the $^{13}$CO data
were kindly provided by J\"urgen Ott), as well as the [C\,{\sc ii}] 158 $\mu$m
line observed spectrally resolved by \cite{Boreiko91} to better constrain the
physical parameters of the emitting gas.   The $^{12}$CO and
  $^{13}$CO $J=1\to 0$ lines  have been observed
with the ATNF Mopra 22\,m telescope fully-sampled with its 33\arcsec\,beam at
115 GHz and later convolved to 45\arcsec\,resolution, and the intensities are
converted to the main beam brightness temperature scale using
$\eta_{\mathrm{mb}}=0.42$.  The [C\,{\sc ii}] 158 $\mu$m line has been
observed with the Kuiper Airborne Observatory with its 43\arcsec\,beam
\citep{Boreiko91} at 158\,$\mu$m and its intensity is quoted on a main beam
brightness temperature equivalent intensity scale. The pointing of the KAO
observations is about $25$\arcsec\,further south than the NANTEN2 position.
We take their values without further corrections. In fact, the integrated
intensity of the [C\,{\sc ii}] line from the KAO observations of $4.3 \times
10^{-4}$ erg cm$^{-2}$ s$^{-1}$ sr$^{-1}$ is consistent within 20\% with the
integrated intensity read off the [C\,{\sc ii}] map published by
\cite{Israel96}, observed velocity unresolved with the FIFI spectral imaging
instruments on the KAO.

To estimate adequate errors for the line intensity ratios, we assume that the
absolute calibration of all the independently observed lines is about 20\%.
Table~\ref{tab:gauss_corrected} quotes the thus derived total errors, obtained
by quadratically summing up the formal fit error and the calibration
uncertainty.

\begin{table*} [t]
\caption{Observed line  integrated intensity ratios, corrected for beam coupling to an 
assumed common  
source size of FWHM of
  60\arcsec }
\label{tab:gauss_corrected}
\centering
\begin{tabular}{c c c c c}
\hline\hline
Species & Ratio$^{(a)}$ & fit error & calibration error &
total error \\
\hline
\\[-0.4cm]
\\
$^{12}$CO $J = 7 \to 6$ / CO $J = 4 \to 3$ 
& 0.30 & 0.008 & 10\% & 0.031 \\

[C\,{\sc i}] $^3$P$_2\to ^3$P$_1$  / [C\,{\sc i}] $^3$P$_1\to ^3$P$_0$  
& 1.05 &0.06&10\%& 0.12 \\
$^{13}$CO $J = 4 \to 3$ / $^{12}$CO $J = 4 \to 3$ 
& 0.10 &0.011& 10\%& 0.015 \\

[C\,{\sc i}] $^3$P$_2\to ^3$P$_1$  / CO $J = 7 \to 6$ 
& 0.56 &0.052&  & 0.052 \\

$^{12}$CO $J=1\to 0$ / $^{12}$CO $J=4\to 3$
& 0.69 &0.10& 20\% & 0.17 \\
$^{13}$CO$J=1\to 0$ / $^{12}$CO$J=1\to 0$ 
& 0.12 &0.02& 20\% & 0.03 \\

[C\,{\sc ii}] / $^{12}$CO $J=4\to 3$
& 1.1 & 0.1 & 20\%  & 0.24\\
\hline
\end{tabular}
\end{table*}

\section{Escape probability radiative transfer: excitation analysis}
\label{sec:escape-prob-radi}

We now use the observed line intensity ratios and absolute intensities
to constrain the physical properties of the emitting gas. We start by
using only the intensity ratios, which are independent of the beam
filling of the source.  We analyze separately the CO emission and the
atomic carbon emission, but assume that both originate from the same
material in the beam with identical beam filling factors. This is
motivated by the fact that the beam sizes at 460-490 and 810 GHz at
the distance of the LMC corresponds to linear scales of 10 and 7 pc,
respectively.  Moreover, both CO and [C\,{\sc i}] emission arise from
dense molecular material, more specifically from the UV illuminated
surface of cloud structures unresolved at this linear scale.

We use the escape probability radiative transfer model by \cite{Stutzki85} to
calculate the line intensities as a function of kinetic temperature $T_{kin}$,
cloud density $n_{H2}$ and column density per velocity interval $N/\Delta$v of
the species of interest (note that the column density parameter in the
radiative transfer model is the radial column density of the spherical clump;
the beam averaged column density in a beam larger than the clump thus is a
factor of
$\frac{1}{r}\,\left(\frac{4\pi\,r^3}{3}\right)/\left(\pi\,r^2\right)=
\frac{4}{3}$ higher).  The collision rate coefficients used are from
\cite{Flower1985} for CO-H$_2$  and from \cite{Schroeder1991} for 
  C$^0$-H$_2$.  The escape probability model produces peak brightness
temperatures of an individual spherical model clump, averaged over the
projected area of the clump on the sky. To compare with the observable
brightness of the model clump, this has to be scaled down by the beam filling
factor of the individual clump. Within the standard model of a molecular cloud
consistent of many fragments which move around with a virialized velocity
dispersion, the ratio of the observed to the model line intensity, once
corrected by the beam filling factor, is thus given by the ratio between the
clump internal linewidth and virial velocity dispersion.  For optically thin
emission of the individual clumps, and also for moderately optically thick
clumps, the observed/model intensity ratio gives the ratio of beam averaged
column density to clump intrinsic column density, as long as the total
integrated line intensity still increases linearly with the number of clumps
(no velocity crowding).  We will use this below, when we compare the absolute
line intensities of $^{12}$CO, $^{13}$CO and [C\,{\sc i}] with the model
predicted intensities.

\subsection{[C\,{\sc i}]-emission}
\begin{figure}[t]
  \centering
  \includegraphics[width=0.38\textwidth,angle=0]
{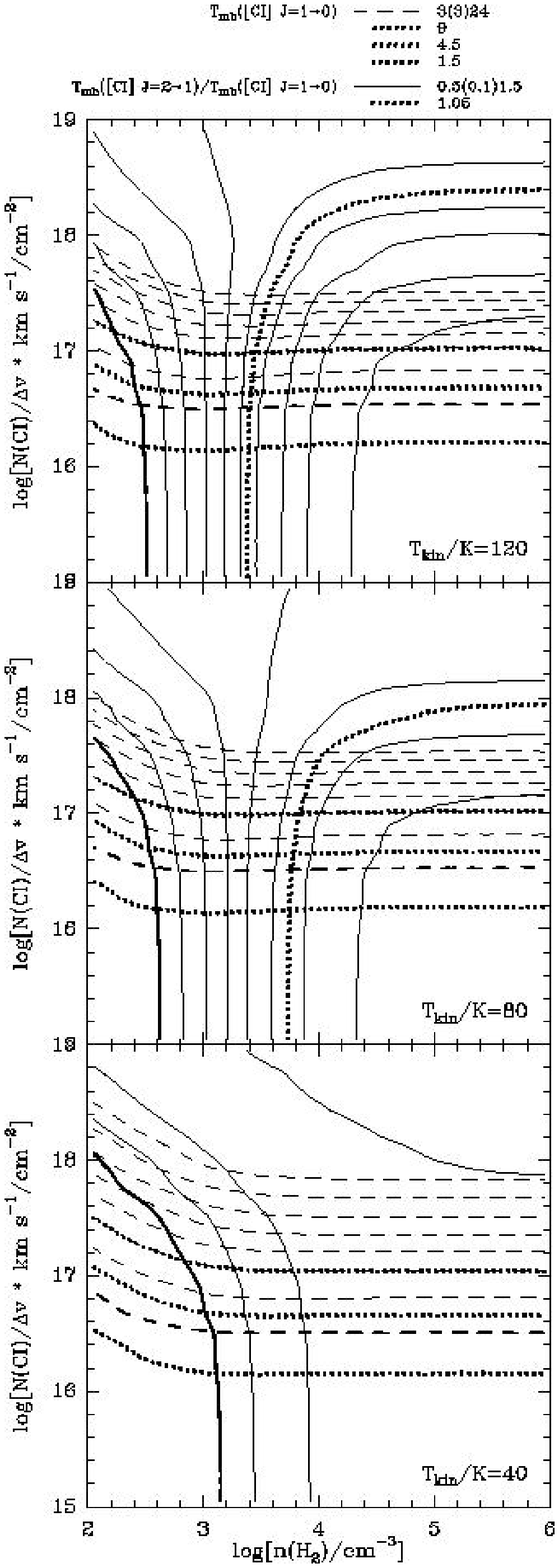}
     \caption{Comparison between observed and escape probability model
       predicted [C\,{\sc i}] line ratios 
       (dashed contours)
       and absolute intensity of the 
       [C\,{\sc i}] $^3$P$_1\to ^3$P$_0$ line 
       (solid contours). 
       Contour levels are given at the top in the
       format ``min(step)max'', with the minimum contour level plotted
       as a thicker line.
       The levels of the dotted contours are the observed values. For the 
       absolute intensity contours these are shown 
       additionally scaled by a beam filling
       factor of 1/3 to 1/6. 
 }
\label{fig:ci_ratio}
\end{figure}

The observed line intensity ratio of the two [C\,{\sc i}] lines of about 1.05
(Table~\ref{tab:gauss_corrected}) is compared with the escape probability
radiative transfer prediction in Fig.~\ref{fig:ci_ratio}.  We present model
line ratios, in a stack of 3 plots corresponding to kinetic temperatures of
40, 80, and 120 K, as a function of density and  radial column density per
  velocity interval, covering the range from 10$^2$ to 10$^6$ cm$^{-3}$ in
density, and 10$^{15}$ to 10$^{19}$ cm$^{-2}$ / (km\,s$^{-1}$) in column
density per velocity interval\footnote{Note that the beam average column
  density per velocity interval of an unresolved clump, as explained above, is
  a factor of 4/3 larger.}.  The line ratios and absolute intensities
discussed further down are presented in the same way and cover the same
parameter range.  Clearly, the observed value can only be reproduced by the
model at moderately high temperatures of above about 60 K, and at densities of
the [C\,{\sc i}]-emitting material of at least $2\times10^3$ cm$^{-3}$ at high
temperatures, and slightly higher, around $6\times10^{3}$ cm$^{-3}$ at the
lower limit of the temperature range.  At these densities, slightly higher
than the critical densities of the atomic carbon fine structure lines, the
carbon lines should be close to thermalized.  Fig.~\ref{fig:ci_ratio} also
shows the model predicted absolute intensity of the [C\,{\sc i}] $^3$P$_1\to
^3$P$_0$ line.  The analysis of the different CO-lines presented below argues
for a beam filling factor of about 1/6, so that the observed peak brightness
of about 1.5\,K should be scaled up to an intrinsic brightness of about 9\,K
corresponding to a C\,{\sc i} column density model parameter of $10^{17}$
cm$^{-2}$/(km\,s$^{-1}$).  In comparison to the kinetic temperature range
derived above and together with the densities inferred, we conclude that the
low observed and beam filling corrected brightness implies that the [C\,{\sc
  i}] lines are well in the optically thin regime. Thus, multiplication by the
inferred beam filling factor is appropriate. The intersection of the line
ratio and absolute line intensity curves, lying on the vertical branch of the
[C\,{\sc i}] line ratio curves in the density/column density plane,
independent of the value of the beam filling factor, implies that the density
of the emitting material is in fact close to the lower limit values quoted
above.

\subsection{CO-emission}
\label{sec:co-emission}

\begin{figure}[t]
  \centering
  \includegraphics[width=0.38\textwidth,angle=0]
{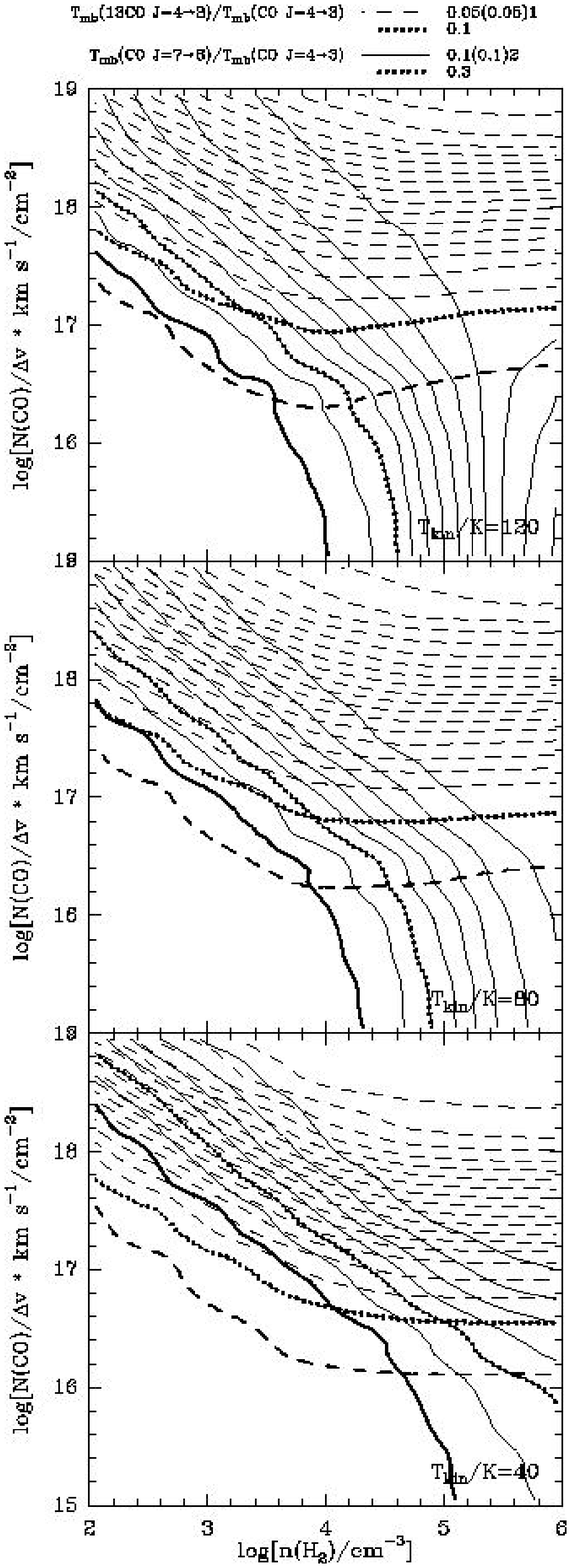}
     \caption{Comparison between observed and escape probability model
       predicted $^{12}$CO $J = 7 \to 6$ to $J = 4 \to 3$ (solid contours) and
       $^{13}$CO to $^{12}$CO $J =4 \to 3$ (dashed contours) line ratios. The
       plots assume a fractional abundance of $^{13}$CO/$^{12}$CO of 40.}
\label{fig:co_ratios}
\end{figure}

>From the NANTEN2 data we obtain two CO line intensity ratios, namely
${T_\mathrm{mb}({^{12}\mathrm{CO}}~J =7\to
  6)}/{T_\mathrm{mb}({^{12}\mathrm{CO}} ~J=4\to 3)}$ and the isotopomeric
ratio ${T_\mathrm{mb}(^{13}\mathrm{CO}~J=4\to
  3)}/{T_\mathrm{mb}(^{12}\mathrm{CO}~ J=4\to 3)}$.  The predicted line
intensity ratios are plotted in Fig.~\ref{fig:co_ratios}.  This plot (and the
following ones) assume a fractional abundance of $^{13}$CO/$^{12}$CO of 40 by
shifting the contours for the $^{^13}$CO lines accordingly along the column
density axis.  A value for the $^{13}$CO/$^{12}$CO abundance ratio in the
N159W region of 40 is cited by \cite{Johansson94} and \cite{Israel2003} quote
a value of 50, whereas \cite{Heikkilae99} derive a substantially lower value
of only 21. Their abundance ratios for other $^{12}$C and $^{13}$C carriers,
namely H$^{13}$CO$^{+}$ and $^{13}$CS, are higher. This may indicate, that the
isotopomeric CO ratio is indeed affected by fractionation, whereas the
elemental $^{12}$C/$^{13}$C ratio is around 40.  Fractionation may occur due
to the exchange reaction $^{13}$C$^+$ $+$ $^{12}$CO $\rightleftharpoons$
$^{12}$C$^+$ $+$ $^{13}$CO $+$ 35~K being exothermal and thus preferentially
incorporating $^{13}$C into the molecular form. Selective photodissociation of
$^{13}$CO due to the lower self-shielding would cause the opposite. PDR model
calculations show that the proportion of these effects strongly depend on the
temperature derived for the C$^+$/C\,{\sc i}/CO transition zone and the
comparison between different models gives non-conclusive answers.  If we
constrain the density to values compatible with those derived from the
[C\,{\sc i}]-analysis above, the two observed CO line ratios can only be
reproduced by the model at a temperature of about 80~K.  The assumption of
equal densities is fully justified, because their is no physically plausible
scenario that has CO at low densities of about 100 cm$^{-3}$, while the
C\,{\sc i} is located at higher densities of close to 10$^4$ cm$^{-3}$, as
long as we assume that the lines originate in a PDR-like structure. Note that
the reverse, i.e.\ [C\,{\sc i}] originating from much lower density material
than CO, namely from a more diffuse halo of dense molecular clouds, would be
also reasonable.

The intersection of the two observed CO-line ratios constrains the CO-column
density model parameter to a value of about 10$^{17}$cm$^{-2}$/km\,s$^{-1}$ at
80\,K.  Recall that we show the plots for an abundance of $^{13}$CO/$^{12}$CO
of 40.  The column density thus derived changes roughly in proportion to the
assumed isotopomeric abundance ratio, as is confirmed by running the model for
different values; we do not show the corresponding figures here, in order to
save space.

\begin{figure}[t]
  \centering
  \includegraphics[width=0.38\textwidth,angle=0]
{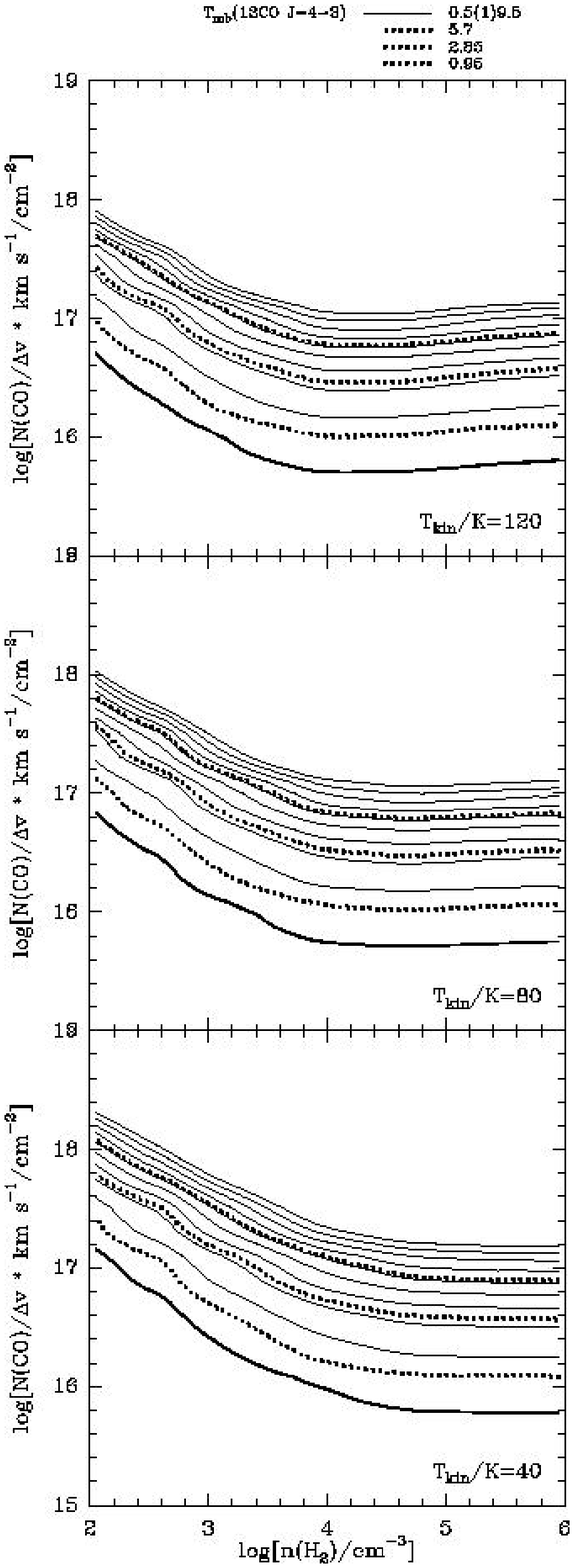}
     \caption{Escape probability model predictions and the observed absolute
       intensity of $^{13}$CO $J=4\to 3$.  The three dotted contours show the
       observed value and the observed intensities scaled up by a factor of 3
       and 6 respectively, to match the column density derived from the line
       ratios at the temperature of about 40~K, respectively 80~K.}
\label{fig:13co_abs}
\end{figure}

Fig.~\ref{fig:13co_abs} shows that the model predicted clump averaged absolute
line intensity of the $^{13}$CO$J=4\to 3$ line at the column densities and in
the density regime constrained by the observed line ratios as discussed above,
is about 6 times higher than the observed one at a temperatures of about
80\,K.  This implies a beam filling factor of the CO material of about 1/6.

\begin{figure}[t]
  \centering
  \includegraphics[width=0.38\textwidth,angle=0]
{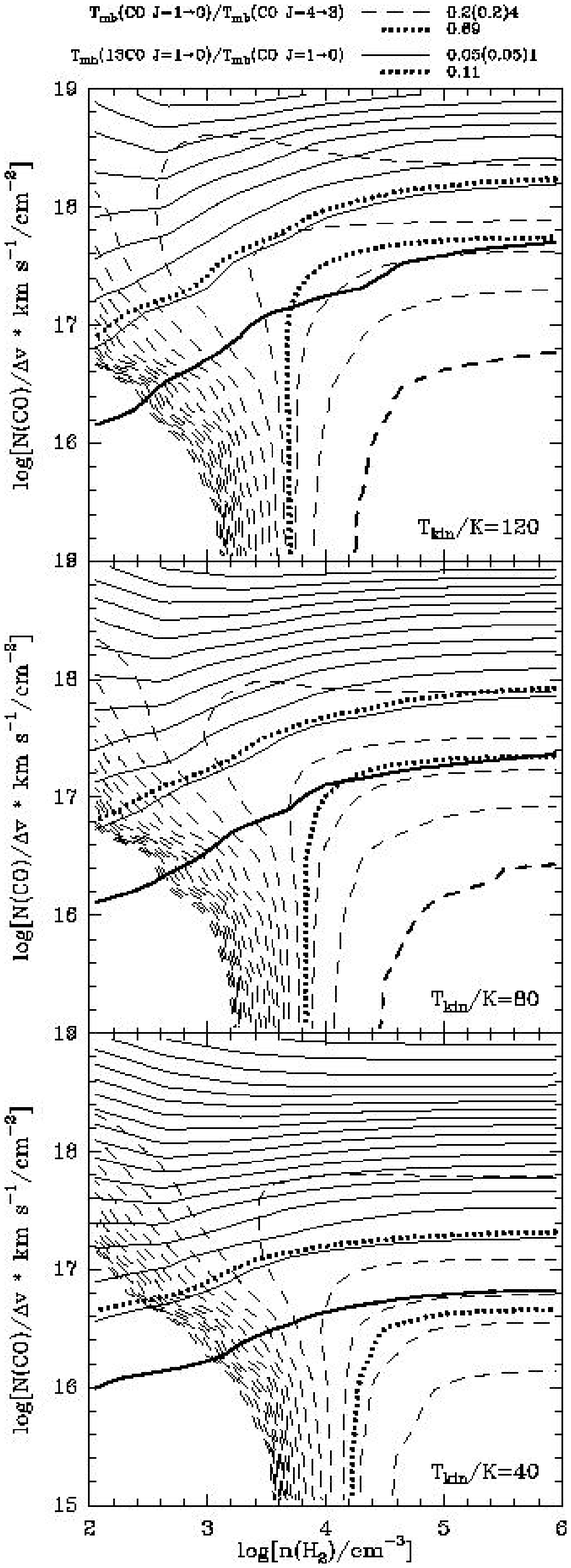}
     \caption{comparison between observed and escape probability model
       predicted $^{12}$CO $J = 1 \to 0$ and $J = 4 \to 3$ line ratios (dashed
       contour) and the $^{13}$CO to $^{12}$CO $J=1\to 0$ line ratio (solid
       contour).  }
\label{fig:co_ratio_astro}
\end{figure}

We now consider in addition the line intensity ratios involving the
$^{12}$CO and $^{13}$CO $J=1\to 0$ transition.  The line intensity
ratio between $^{12}$CO $J=1\to 0$ and $4\to 3$ of 0.69 traverses the
density/column density regime of $n=10^4$ cm$^{-3}$ and about
$10^{17}$ cm$^{-2}$/ (km\,s$^{-1}$) derived from the other CO line
ratios only at temperatures of above around 80\,K and higher (see
Fig.~\ref{fig:co_ratio_astro}). The isotopic $J=1\to 0$ line ratio
does not traverse this regime at all.  It always requires substantially
higher column densities than allowed by the line ratios of the
higher$-J$ lines.  This implies that reducing the $^{12}$CO/$^{13}$CO
abundance ratio from the value of 40 assumed here to lower values,
will bring the $J=1\to 0$ ratio into better agreement, although
Fig.~\ref{fig:co_ratio_astro} shows that lowering it by just a factor
of 2 down to about 20 would not be sufficient. We report (without
showing a figure), that the agreement becomes reasonably good at lower
temperatures of 20 to 40\,K, indicating that a temperature gradient
with the lower$-J$ CO emission originating in lower temperature gas and
showing stronger isotopic fractionation can give a consistent picture,
thus confirming the low $^{12}$CO/$^{13}$CO ratio of about 20 reported
by \cite{Heikkilae99}.

To summarize, the escape probability radiative transfer analysis implies that
the CO and $^{^{13}}$CO emission observed in N159W, if it originates in gas
with densities close to 10$^4$ cm$^{-3}$ as constrained by the [C\,{\sc i}]
emission, comes from gas with kinetic temperatures of about 80\,K and with a
source intrinsic total column density of $10^{17}$
cm$^{-2}$/(km\,s$^{-1}$) and a filling factor of about 1/6. 
The $^{12}$CO $J=7\to 6$, $J = 4\to 3$, $J = 1\to 0$ and the $^{13}$CO
$J = 4\to 3$ and $J = 1\to 0$ line ratios taken together 
are not consistent with a single temperature
component, but require a range of temperatures 
and possibly isotopic fractionation.

\subsection{CO  and  [C\,{\sc i}] column densities}

The escape probability radiative transfer analysis gives column densities for
both CO and C\,{\sc i} of 10$^{17}$ cm$^{-2}$/km\,s$^{-1}$.  But the $^{12}$CO
and $^{13}$CO $J = 1\to 0$ lines indicate an additional, cooler component.
The derived column density ratio [C\,{\sc i}] / CO of 1.0 thus is an upper
limit due to the unknown contribution of the lower temperature CO-material.

As discussed above, the beam average column density is a factor of 4/3 higher
than the column density model parameter.  Together with the observed line
width of 7.3 km\,s$^{-1}$ and the beam filling factor of 1/6, both CO and
C\,{\sc i} thus have a beam average column density of $ 1.6 \times 10^{17}$
cm$^{-2}$. Assuming the canonical abundance CO/H$_2$ of $8 \times 10^{-5}$ and
the same for C\,{\sc i}, both tracers correspond to a mass in the beam of 3000
M$_\odot$ each, and 6000 M$_\odot$ in total. Considering that our beamsize of
38\arcsec\,FWHM corresponds to 9.5 pc at the distance of the LMC and hence
covers only part of the N159W cloud's extent of about 15 pc, this is in good
agreement with the CO-luminosity mass of $1.9 \times 10^4$ M$_\odot$ derived
by \cite{Johansson98}.

\section{PDR-model analysis}
\label{sec:pdr-model-analysis}

In the previous section we have analyzed the excitation conditions of both
[C\,{\sc i}] and CO/$^{13}$CO separately. These conditions correspond to
  typical values of the line-emitting material. In the following, we
  investigate whether the observations can be explained in terms of PDRs
  distributed in a clumpy interstellar medium.   The modest temperatures,
with a tendency towards lower temperatures for the CO lines than for the
[C\,{\sc i}] lines, are roughly consistent with those expected for the mid$-J$
CO and [C\,{\sc i}]-emitting layers in a PDR illuminated by the modest
UV-field of $\chi$ about 175, independently derived for the LMC N159W region
from the FIR continuum luminosity \citep{Israel96}.  Compared to typical star
forming regions in the Milky Way, the estimated column density ratio of
C\,{\sc i} relative to CO of up to unity is rather high and this is expected
to be the case in a lower metallicity environment such as the LMC with z=0.4.
Similarly, the [C\,{\sc ii}] emission is expected to be stronger due to a
larger extent of the C-ionized PDR surface layer resulting from the weaker UV
attenuation by dust and weaker self shielding of CO at low metallicities.

\cite{Bolatto99}, in a simple semi-analytical model (their model A) predict
this behavior for the carbon species with the assumption of finite size
clumps making up the emitting gas; the PDR depth scale is determined by the UV
absorption per nucleon, which is due to heavy elements (in the form of dust,
carbon etc.) and thus, referred to the hydrogen column density, scales up
inversely with the metallicity. The [C\,{\sc ii}] and [C\,{\sc i}] emitting
layers thus grow in size inversely with metallicity and the molecular core of
the clump, where the CO emission originates, shrinks.

The KOSMA-$\tau$ PDR model \citep{Stoerzer96} allows to self-consistently
calculate both the chemical and temperature structure of spherical clumps and
confirms these basic trends with metallicity \citep{Roellig06}, but takes into
account all details such as the variation of the temperature in the emitting
region of each species with varying metallicity. In order to model the
emission from a cloud complex such as the N159W region, one additionally has
to take into account the fact, that the emission is unlikely to arise in a
single clump, but that the emitting complex shows structure on all scales.
Quantitative studies of the structural properties in many star forming regions
in the Milky Way
\citep[e.g.][]{Williams94,ElmegreenFalgalore96,Stutzki88,Kramer98,Heithausen98,Heyer98},
but also on larger scales in the Milky Way and external galaxies
\citep{Staminirivic2000,Staminirovic2001,Fukui01,Kim2007}, show that the
observed velocity channel maps can be decomposed into clumps with a power law
mass spectrum $dN/dM \propto M^{-\alpha}$ with a clump mass spectral index of
1.8 to 1.9; without the velocity dispersion due to turbulence, this would not
be possible, as in many cases the clumps overlap along the line-of-sight and
can only be separated by the additional velocity information. An independent
approach is to study the power spectra of the observed maps, either directly
\citep[e.g.][]{ElmegreenFalgalore96} or via the $\Delta$-variance analysis
\citep{Stutzki98}, which shows that these often follow a power law.
\cite{Stutzki98} showed, that both approaches are consistent with the
additional assumption of a power-law mass-size relation for the individual
clumps, $M \propto r^{\gamma}$, with an index $\gamma$ of about 2.3. Note that
values of $\gamma > 2$ imply increasing density with decreasing clump size,
i.e. denser small scale structures, as expected for a turbulent medium, where
the small scale structures are created by local compression due to transient
waves; $\gamma=2$ would imply constant column density of the structures,
independent of size.  Although the demonstration of a clear mass-size power
law and the derivation of the exact power exponent is rather difficult and
uncertain, in the few cases where it has been possible it confirms values
around the one quoted above; an example is the analysis of the Polaris flare
region \citep{Heithausen98}.

Based on this scenario, an obvious approach is to model the observed emission
of a cloud complex by combining the emission of many clumps, individually
modeled with a clump PDR-model such as KOSMA-$\tau$, which follow a power law
mass spectrum and mass-size relation. This approach has been successfully
applied to several individual regions
\citep{Kramer2008,Mookerjea2006,Kramer2004,Kramer2005} , but also to the
global emission of the Milky Way in the submm/FIR cooling lines, as observed
by COBE \citep{Cubick2007}.  We will use it here to analyze the combined set
of observed lines in LMC N159W. The details of the model are explained in
\citet{Cubick2007}.

Fixing the clump mass spectral index and the mass-size power-law index
to $\alpha=1.8$ and $\gamma=2.3$ respectively, the clumpy cloud
complex is characterized by the mean density of the total clump volume
(each clump being modeled as a Bonnor-Ebert sphere with $\propto
r^{-1.5}$ density law, so that the clump average density is twice the
surface density), the total mass of the clump ensemble, and the upper
and lower mass cut-off of the mass spectrum. As long as the individual
clumps are not too densely packed spatially and in velocity, so that
their emission does not overlap along the line-of-sight and in
frequency, changing the total clump mass simply changes the absolute
brightness of the clump ensemble by adding or removing material with
identical mass distribution; equivalently, it changes the area and
volume filling factor of the clump ensemble. In the model fit, it is
thus used as the parameter that adjusts to the absolute intensity
observed and it can be ignored, as long as one considers only relative
intensities.

Keeping the other parameters fixed, but changing the upper mass cut-off shifts
the distribution of the material to larger or smaller clumps and thus will
affect the relative intensities of the various lines according to their
preferential origin from the clump cores or surfaces. Lowering the lower mass
cut-off simply adds more and more smaller clumps, which, as long as the lower
mass cut-off is sufficiently low, are small enough to correspond basically to
PDR-surface layers; given the above value of the mass-spectrum index, the
small clumps do not contribute to the total mass, i.e. amount of material, and
thus this change does not significantly affect the observed emission. A
somewhat open issue, discussed in more detail in \citet{Cubick2007} concerns
the fact, that one can argue about the stability and life-time of such small
clumps and to what extent the mid$-J$ CO emission from these small and dense
structures affects the overall emission of the ensemble in these lines. We do
not discuss this issue further here. For the models discussed below, the low
mass cut-off was fixed to $10^{-3}$ $M_\odot$.

The free model parameters are thus, as long as we consider only relative
intensities, the UV-field, parametrized by $\chi$ in units of the Draine
field, the clump-volume mean density $n$, the metallicity $z$ and the upper
mass cut-off $M_{max}$. For the following analysis, we first fix the
metallicity to the canonical value for the LMC of $z=0.4$; later on we
discuss, how the results vary with metallicity. We then do a least-square fit
in the $\chi$, $n$ -parameter plane for different values of $M_{max}$. The
least-square fit analysis includes all line ratios listed in
Table~\ref{tab:gauss_corrected}, i.e. the relative intensities of the full
suite of low$-J$ and mid$-J$\ CO and $^{13}$CO lines, both [C\,{\sc i}] lines,
and the [C\,{\sc ii}] line.  Unfortunately, the PDR model grid is presently
only available for a $^{12}$C/$^{13}$C elemental abundance of 60, slightly
higher than the experimentally determined value for the LMC N159W cloud.

\begin{figure}[h]
  \centering
  \includegraphics[width=0.5\textwidth,angle=0]{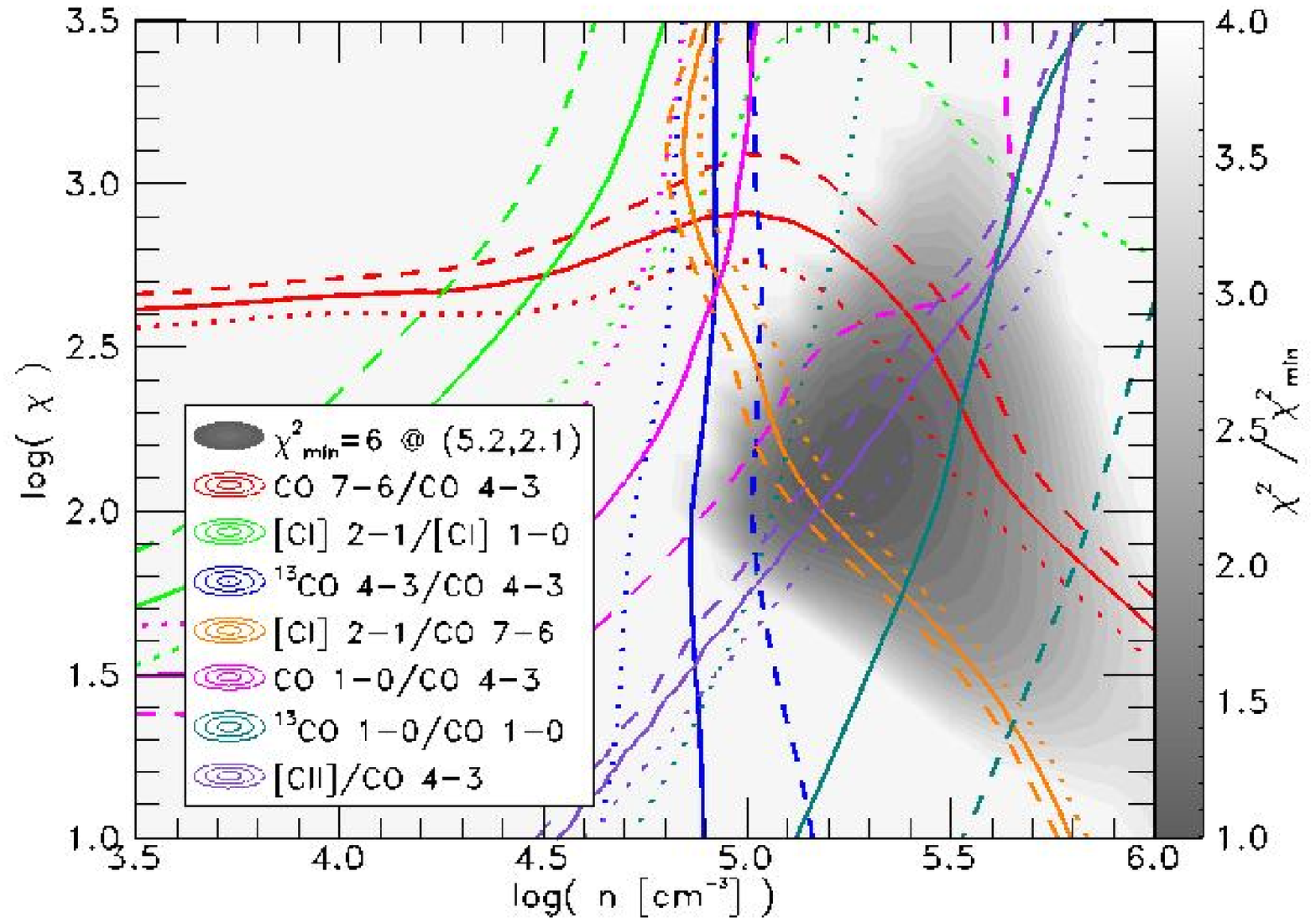}
\caption{
  Clumpy KOSMA-$\tau$ PDR-model fit to the observed line ratios overlaid onto
  the gray shaded contours of the corresponding reduced $\chi^2$. Note that
  the observed line ratios are plotted as solid lines, with the lower and
  upper limits as dotted respectively dashed lines. The $\chi^2$ gray-scale
  ranges from one to four times the minimum $\chi^2$-value of 6 at a density
  of $\log(n\, [\rm{cm}^{-3}]) = 5.2$ and an incident FUV-flux of $\log(\chi)
  = 2.1$.  }
\label{fig:plot_pdr_fit}
\end{figure}

We get the best fit, with a normalized $\chi^2_{min}=6$ for a UV-field of
$\chi$ around 220 and a density $n$ of a few times $10^5$ cm$^{-3}$. This is
obtained at a rather low upper-mass cut-off of 0.1 $M_\odot$.
Fig.\ref{fig:plot_pdr_fit} shows the $\chi^2$-distribution in the parameter
plane and the contours of the observed line ratios overlayed.  Considering
that we fit seven line ratios with 2 parameters, this is an excellent fit.

\begin{figure}[t]
  \centering
  \includegraphics[width=0.5\textwidth,angle=0]{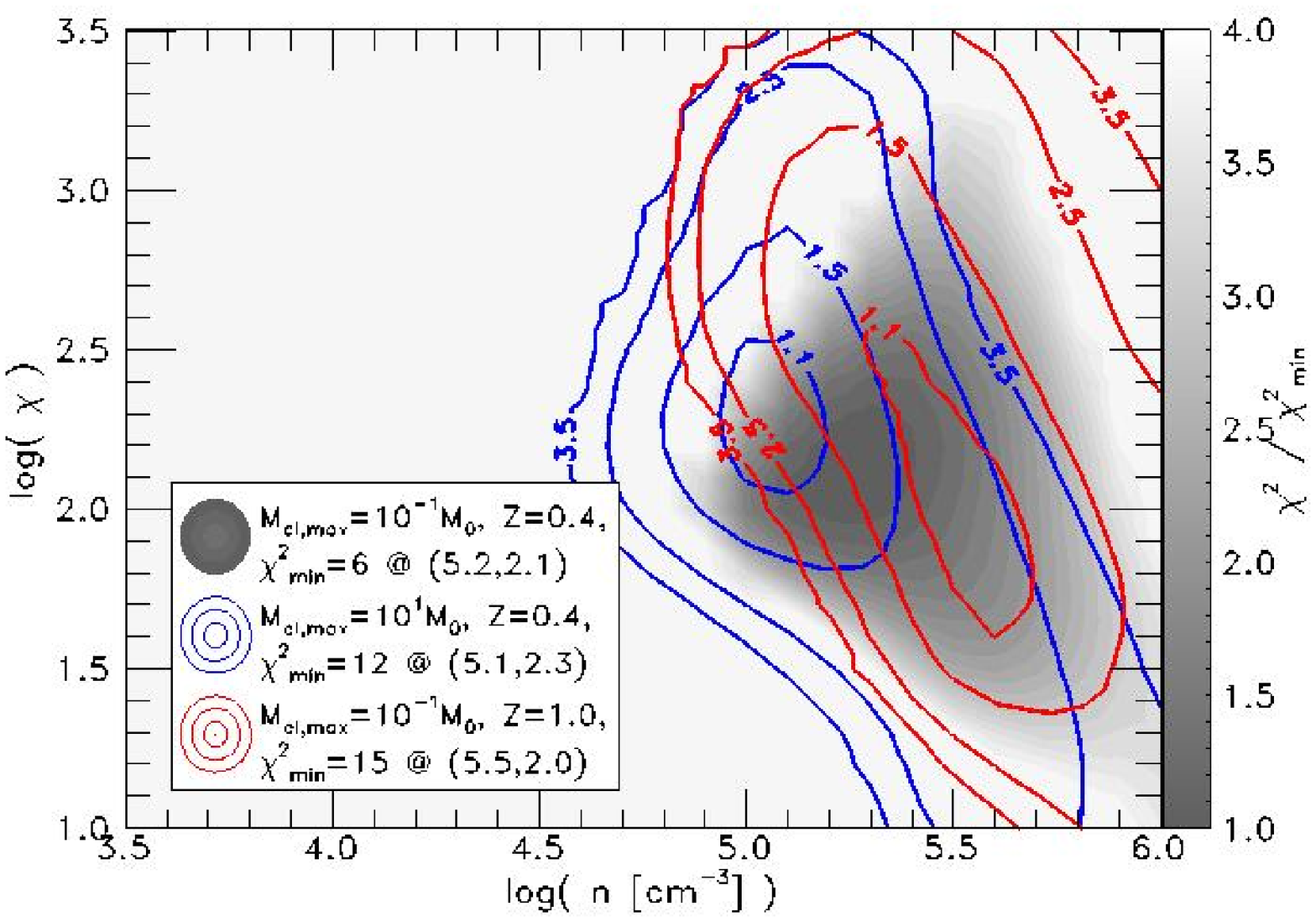}
\caption{
Reduced $\chi^2$ contours of the clumpy KOSMA-$\tau$ PDR-model results
for different values of upper clump mass limit $M_{cl,max}$ of the clump 
ensemble and metallicity $Z$.
Note that all contours are given in values of the individual minimum 
$\chi^2$ at the corresponding values of the density and FUV-flux 
inserted in the legend.
} 
\label{fig:plot_pdr_fit_var}
\end{figure}

We next discuss the dependency of the best fit parameters on varying the upper
mass cutoff and the metallicity. Representative $\chi^2$-contour plots are
shown in Fig.\ref{fig:plot_pdr_fit_var}.  Increasing the upper-mass cut-off to
1 M$_\odot$ and 10 M$_\odot$ increases the $\chi^2_{min}$ to a value of 7.5
and 12.8. This shallow rise of less than a factor of 2 in $\chi^2$ per decade
in $M_{max}$, shows that the upper mass-cutoff is not very stringently
constrained, but indicates that the clumpy cloud PDR scenario requests the
emitting gas to show lots of surface and to be broken up into small scale
structures.  Varying the metallicity to the solar value of 1, the
$\chi^2_{min}$ is 15 and 22 respectively for a $M_{max}$ of 0.1 and 10
M$_\odot$, so that the PDR model clearly favors the canonical metallicity of
the LMC in order to reproduce the observed intensity ratios.
Fig.\ref{fig:plot_pdr_fit_var} shows that the discussed variations do not much
alter the best fit value for the UV-intensity and the density.

Considering the uncertainties about the carbon isotope elemental abundances to
use for the LMC and the open issue of isotope selective fractionation in PDRs,
it is worthwhile to consider a PDR-model fit excluding the $^{13}$C-carrying
species. Without showing the corresponding plot, we report that the best-fit
density and UV-intensity are unaffected; this is due to the fact, that the
line ratios involving the $^{13}$C-carrying species vary rather shallow across
the parameter region of interest. A close inspection of
Fig.\ref{fig:plot_pdr_fit} in fact shows, that neglecting the observed
[C\,{\sc ii}]/CO $J=4\to 3$ ratio in determining the $\chi^2$ -minimum would
allow a solution at slightly lower densities of about $10^5$ cm$^{-3}$ and
higher UV-intensities of slightly above $\chi$=1000, where all other lines
intersect.

Whereas the value for the UV-intensity is consistent with that independently
derived from the FIR luminosity, as discussed in the introduction, and thus
only confirms this result independently, the density regime favored by the
clumpy-cloud PDR-model, $n=10^5$ cm$^{-3}$ or slightly above, is clearly much
higher than other estimates of the density in the LMC N159W region.  Together
with the small value for the upper-mass cut-off this argues for a very clumpy
structure with high density clumps to explain the observed line ratios.

In the last step of the analysis, we can now adjust the predicted absolute
line intensities to the observed value, by changing the total mass of the
ensemble to fit the absolute intensity of e.g. the $^{12}$CO $J=4\to 3$ line.
This gives a mass included in the beam of $2.8 \times 10^4$ M$_\odot$, in good
agreement with the CO-luminosity mass quoted by \cite{Johansson98} and the
mass derived above from the escape probability analysis.  Note that the
clumpy-PDR model fits all lines simultaneously, where as the low$-J$ CO line
analysis is biased towards the lower excitation material and the escape
probability radiative transfer analysis above emphasized the warm and dense gas
traced by the mid$-J$ CO transitions and the [C\,{\sc i}] fine structure lines.

We conclude that the emission of all observed CO lines, from $J=1\to 0$ to
$7\to 6$ for $^{12}$CO, $J=1\to 0$ and $4\to 3$ for $^{13}$CO and both
[C\,{\sc i}] and [C\,{\sc ii}] fine structure lines, is consistently modeled
as originating from a dense, highly clumped medium illuminated by an FUV-field
of $\chi \approx 220$ in the LMC N159W complex. The model fit constrains the
densities to values of about $10^5$ cm$^{-3}$ and preferences the canonical
metallicity of the LMC of about $z=0.4$.

\section{Discussion}
\label{sec:Discussion}

 Previous CO excitation analyses suggest a low-temperature ($T =
  10-20$\,K) molecular gas in N159W \citep{Johansson98,Bolatto05}.  However,
  these studies are only based on CO observations which trace more shielded
  material in the PDR. The excitation analysis using the [C\,{\sc i}] fine
  structure line intensity ratio allow us to better explore the conditions at
  the cloud surfaces, as this ratio is a very good tracer of the surface
  temperature \citep[e.g.][]{Stutzki1997}.  \citet{Bolatto05} and
  \citet{Minamidani2007} derived larger temperatures for this region ($T =
  150$\,K and $T > 30$\,K, respectively).  Our work confirms a high density of
  the bulk of the CO-traced molecular material in the N159W region.  Wide
  ranges of possible gas densities have been suggested by \citet{Johansson98}
  and \citet{Minamidani2007} (10$^{3}-10^{5}$ cm$^{-3}$). Although poorly
  constrained, \citet{Bolatto05} also proposed a two-component solution of low
  density ($n=10^2$\,cm$^{-3}$), high temperature ($T = 100$\,K) and high
  density ($n=10^5$\,cm$^{-3}$), low temperature ($T = 20$\,K).  High
  densities of the more shielded material are also suggested by observations
  of molecular species like CS, SO, and H$_2$CO
  \citep[$n=10^{5}-10^{6}$\,cm$^{-3}$;][]{Heikkilae99}.

We derive a [C\,{\sc i}]/CO column density ratio toward N159W close to 1.0,
possibly slightly lower, depending on the contribution from a cooler CO
component.  This value is large compared with what is typically found in
Galactic dense massive star formation regions (0.04-0.2; for a compilation,
see \citealt{Mookerjea2006}), and therefore argues for a larger C abundance
relative to CO as a result of the enhanced CO photodissociation in
low-metallicity PDRs.   Note that the [C\,{\sc i}]/CO column density ratio
  is not constant in the Galaxy \citep{Mookerjea2006}. High values are also
  found in the less dense outskirts of massive star-forming regions
  \citep{Mookerjea2006} and in high-latitude clouds \citep{Ingals1997}. Recent
  observations have also shown high [C\,{\sc i}]/CO column density ratios in
  extragalactic nuclei \citep{Hitschfeld2007}.   In order to determine the
relative importance of the main carbon species, we derive the C$^+$ column
density, assuming $N_\mathrm{[C\,{\sc II}]}=6.4\times 10 ^{20}$
$I_\mathrm{[C\,{\sc II}]}$\,cm$^{-2}$ (erg s$^{-1}$ cm$^{-2}$
sr$^{-1}$)$^{-1}$ \citep{Crawford1985}\footnote{This is an approximation for
  an excitation temperature of $>91$\,K and a density larger than
  $n_\mathrm{cr}=5 \times 10^{3}$, cm$^{-3}$}, and using the
$I_\mathrm{[C\,{\sc II}]} = 4.3 \times 10^{-4}$ erg s$^{-1}$ cm$^{-2}$
sr$^{-1}$ observed by \citet{Boreiko91}.  With this method, we obtain
$N_\mathrm{[C\,{\sc II}]}=2.75 \times 10^{17}$ cm$^{-2}$.  Using this value
and the CO and CI column densities of $N_\mathrm{CO,C\,{\sc I}}=1.6 \times
10^{17}$ cm$^{-2}$ as derived above, we obtain a relative abundance of the
main carbon species in N159W of $N_\mathrm{[C\,{\sc II}]}:N_\mathrm{[C\,{\sc
    I}]}:N_\mathrm{CO}$=46\%:27\%:27\%.  In Galactic star-forming regions CO
is the most abundant carbon species.  For example, in the DR21(OH)
star-forming region, the relative abundance of the main carbon species is
$N_\mathrm{[C\,{\sc II}]}$:$N_\mathrm{[C\,{\sc
    I}]}:N_\mathrm{CO}$=3\%:10\%:87\% \citep{Jakob2007}.  As we can see, the
abundance of C$^{+}$ and C relative to CO are indeed enhanced in the
low-metallicity gas in N159W.  The different distribution of carbon is
reflected in the enhanced $I_\mathrm{[C\,{\sc II}]}/I_\mathrm{^{12}\mathrm{CO}
  J = 1 \to 0}$ and $I_\mathrm{[C\,{\sc I}]}$/$I_{^{12}\mathrm{CO} J = 1 \to
  0}$ line ratios observed in LMC N159W, relative to Solar metallicity PDRs.
>From the data used in this paper, we obtain $I_\mathrm{[C\,{\sc
    II}]}/I_\mathrm{^{12}\mathrm{CO} J = 1 \to 0}$ $\simeq$ 1.13 and
$I_\mathrm{[C\,{\sc I}] ^3\mathrm{P}_1-^3\mathrm{P}_0 }$/$I_{^{12}\mathrm{CO}
  J = 1 \to 0}$ $\simeq$ 0.22. These values can be compared with the ratios
determined from COBE all-sky measurements \citep{Wright1991} of 0.97 and 0.13,
respectively.

Polycyclic aromatic hydrocarbons (PAHs) are known to have an important
influence in the carbon chemistry, increasing the C and decreasing the CO
abundances in cloud surfaces \citep{LeppDalgarno88,BakesTielens98}.  In order
to assess any effect of PAHs in the PDR modeling, we also compare the
observations with a single-clump version of the KOSMA$-\tau$ model that
includes PAH in the chemical network (see \citet{Pineda07a} for details on the
inclusion of PAHs in the KOSMA$-\tau$ model).  PAHs were also considered in
the PDR model calculations by \citet{Kaufman99}.  We find that the
$I_\mathrm{[C\,{\sc I}]}/I_\mathrm{CO}$ line ratios typically increase by up
to a factor of 10 when PAHs are considered.  As a result, in order to match
the observations, even larger densities than derived above would be required
($n=10^6-10^{7}$\,cm$^{-3}$).  The effects of PAHs in the chemical network of
the clumpy-PDR model will be explored in a future work.  The PAH abundance
might be reduced in low-metallicity molecular gas, as observations revealed
that PAH emission is weak or absent in low metallicity galaxies
\citep[e.g.][]{Madden00,Houck04,Engelbracht05,Madden06}.  However, the
observed PAH emission spectra in the N159-5 H\,{\sc ii} region resembles the
spectra of Galactic H\,{\sc ii} regions \citep{Vermeij02}.

The strength of the FUV radiation field required by the PDR model to explain
the observations ($\chi \simeq 220$) is consistent with the independent
determination made by \citet{Israel96}, using UV, radio continuum, and IRAS
60\,$\mu$m emission ($\chi = 120-350$).  Using the same assumptions made by
\citet{Israel96} to determine $\chi$ from the IRAS 60\,$\mu$m data, but using
the high-angular resolution {\it Spitzer} 70\,$\mu$m map \citep{Meixner06}, we
obtain a consistent value of $\chi = 177$.  This strength of the FUV radiation
field might be considered low for a massive star formation region such as
N159W (for example, in the Orion Bar, \citet{Walmsley00} and
\citet{YoungOwl00} find that $\chi \simeq 10000$).  However, \citet{Israel96}
argue that the low metallicity and low dust-to-gas ratio of this region
increases the FUV photon free path lengths, resulting in a greater geometric
dilution of the FUV radiation field.

\section{Summary and conclusions}

In this paper, we derived the physical conditions of the line-emitting gas in
the N159W region in the LMC. We compared an escape-probability code and a PDR
model with NANTEN2 observations of the $^{12}$CO $J = 4 \to 3$, $J = 7 \to 6$,
and $^{13}$CO $J = 4 \to 3$ rotational and [C\,{\sc i}] $^3$P$_1-^3$P$_0$ and
$^3$P$_2-^3$P$_1$ fine-structure transitions. Our results can be summarized as
follows:

\begin{itemize}
  
\item The analysis of the excitation conditions for both the CO submm-lines
  and the [C\,{\sc i}] fine structure lines shows temperatures of about $T =
  80$\,K and densities of about $n= 10^{4}$ cm$^{-3}$ for the emitting gas.

\item The estimation of the C and CO column densities suggest a high C/CO
  abundance ratio of close to unity compared to that found in Galactic dense
  massive star-forming regions.  This is an indication that CO is more readily
  destroyed in the low-metallicity environment of N159W.  Comparing with an
  estimation of the C$^+$ column density, we confirm that the distribution of
  the main carbon species in PDRs is altered in the low-metallicity molecular
  gas, with larger C$^+$ and C abundances relative to CO. This result is in
  agreement with observations of an enhanced $I_\mathrm{[C\,{\sc
      II}]}/I_\mathrm{CO}$ and $I_\mathrm{[C\,{\sc I}]}$/$I_\mathrm{CO}$
  ratios in N159W, compared with values observed in solar metallicity PDRs.

\item  The PDR modeling, calculating the emission as resulting from an
    ensemble of spherical clumps with a power law mass spectrum and a power
    law mass-size relation, providing a simple representation of the fractal
    structure of the ISM, gives a consistent fit to all observed line ratios
    for a moderate strength of the FUV radiation field of $\sim$220\,$\chi_0$.
    This value is consistent with a previous independent estimation made by
    \citet{Israel96}. It constrains the average density of the clump ensemble
   to about 10$^5$ cm$^{-3}$,  confirming previous indications of
    high-density CO-traced molecular material in the LMC N159W region.  The
  clumpy cloud PDR model also gives a relatively small upper clump mass cutoff
  for the clump ensemble, indicating a very much broken up structure of the
  ISM in N159W.

\end{itemize}

 Similar studies in other regions in the LMC with NANTEN2 combined with
  observations from upcoming observatories such as Herschel and SOFIA, will
  improve our knowledge about the impact of star formation in the
  low-metallicity molecular gas. This will help us to improve our current
  understanding of the processes of star formation in the young Universe. 
\begin{acknowledgements} 

  The NANTEN2 project (southern submillimeter observatory consisting of a
  4-meter telescope) is based on a mutual agreement between Nagoya University
  and The University of Chile and includes member universities from six
  countries, Australia, Republic of Chile, Federal Republic of Germany, Japan,
  Republic of Korea, and Swiss Confederation. We acknowledge that this project
  could be realized by contributions of many Japanese donors and companies.
  This work is financially supported in part by a Grant-in-Aid for Scientific
  Research from the Ministry of Education, Culture, Sports, Science and
  Technology of Japan (No.¥ 15071203) and from JSPS (No.¥ 14102003 and No.¥
  18684003), and by the JSPS core-to-core program (No.¥ 17004).  This work is
  also financially supported in part by the grant SFB\,494 of the Deutsche
  Forschungsgemeinschaft, the Ministerium f\"ur Innovation, Wissenschaft,
  Forschung und Technologie des Landes Nordrhein-Westfalen and through special
  grants of the Universit\"at zu K\"oln and Universit\"at Bonn.  LB and MR
  acknowledges support from FONDAP Center of Excellence 15010003.  We thank
  F.\,Israel for providing some older SEST $^{12}$CO and $^{13}$CO $J=1\to 0$
  data for cross checking the calibration of the MOPRA data.  We also thank
  J\"urgen Ott for providing the Mopra $^{13}$CO data.  Data reduction of the
  spectral line data was done with the {\tt gildas} software package supported
  at IRAM (see {\tt http://www.iram.fr/IRAMFR/GILDAS}). This research has made
  use of NASA's Astrophysics Data System Abstract Service.

\end{acknowledgements}

\bibliographystyle{aa} \bibliography{./papers}

\end{document}